\documentclass[aps,prl,twocolumn]{revtex4-1}
\usepackage{graphicx}
\usepackage{amsmath,amsthm}
\usepackage{dcolumn}
\newcommand{\emone}{^\textnormal{-1}}%
\newcommand{\smone}{_\textnormal{-1}}%
\newcommand{\ezero}{^\textnormal{0}}%
\newcommand{\szero}{_\textnormal{0}}%

\begin{document}


\title{Quantum gravity simulation by non-paraxial nonlinear optics}
\author{Claudio Conti$^{1}$}
\affiliation{
$^1$Institute for Complex Systems ISC-CNR and Department of Physics, University Sapienza, Piazzale Aldo Moro 5, 00185, Rome (IT)
}
\email{claudio.conti@uniroma1.it}
\date{\today}

\begin{abstract}
We show that an analog of the physics at the Planck scale can be found in the propagation of tightly focused laser beams. Various equations that occur in generalized quantum mechanics are formally identical to those describing the nonlinear nonlocal propagation of nonparaxial laser beams. The analysis includes a generalized uncertainty principle and shows that the nonlinear focusing of a light beam with dimensions comparable to the wavelength corresponds to the spontaneous excitation of the so-called maximally localized states. The approach, driven by the ideas of the quantum gravity physics, allows one to predict the existence of self-trapped subwavelength solitary waves for both focusing and defocusing nonlinearities, and opens the way to laboratory simulations of phenomena that have been considered to be inaccessible.
\end{abstract}


\maketitle
A rapidly expanding research direction in theoretical physics concerns the investigation of 
a generalized uncertainty principle (GUP).\cite{Kempf95,Kempf96,Adler99,Akhoury03,Das08,Sailer13}
Quantum gravity (QG) models predict 
fuzziness and discretization in the geometry, which result in the fact that the uncertainty of
the spatial coordinate $\Delta X$ cannot be smaller than a minimal quantity $\Delta X_{min}$.
The simplest GUP \cite{Kempf95} reads as
 \begin{equation}
 \Delta X \Delta P\geq \frac{\hbar}{2} (1+\beta \Delta P^2)\text{,}
 \label{GUP}
 \end{equation}
with $\Delta P$ the momentum uncertainty.  Equation (\ref{GUP}) implies $\Delta X\geq \hbar \sqrt{\beta}$; 
standard quantum mechanics (QM) is retrieved for $\beta=0$.  
Various upper bounds to $\beta$ have been proposed. \cite{Das08,Das09}
Among the many consequences of the GUP, we mention corrections to the black-body radiation spectrum, 
and to the cosmological constant. \cite{Chang02,Trabelsi10} 
Analyses of GUP effects in the mechanical vibration of macroscopic objects have been reported. \cite{Huan13,Pikovski12,Bekenstein12,ACamelia13} Theoretical developments include the existence of a maximal momentum $P$. \cite{Pedram12}
This may be related to another known fundamental limit to $\Delta X$,
from special relativity (SR), given by the reduced Compton wavelength $\lambdabar_C=\hbar/mc$ \cite{LandauBookQED},
with $m$ the particle rest mass and $c$ the vacuum light velocity.
$\lambdabar_C$ is much larger than the Planck scale  $\l_P\cong 10^{-35}$~m, which 
is typically retained as $\Delta X_{min}$ in QG ($\lambdabar_C\cong 10^{-12}$~m for the electron).

In several respects, the analysis of nonlinear wave equations with GUP is un-explored. In QG nonlinear effects are expected to be extremely relevant because of the energy levels needed to access the Planck scale, and because of the nonlinearity due to background independence. Nonlinear mechanisms are also known to be due to the particle-anti-particle production expected at the scale of the Compton wavelength \cite{LandauBookQED}, and nonlinear
modifications to Maxwell equations have been predicted in loop quantum gravity \cite{Gambini99,Alfaro02}.
In condensed matter physics, nonlinearity is due to the atom-atom interaction in ultra-cold gases and Bose-Einstein condensates,
where investigations of the effects of Planck scale physics have been also reported \cite{Mercati10,Briscese12,Castellanos13}.

\noindent Here we show that all the mentioned theoretical developments naturally apply to the description of 
tightly focused laser beams\cite{Kempf00} with size smaller than the wavelength and propagating
in local and nonlocal nonlinear media.\cite{ Granot99, Ciattoni:05}
This analogy allows to obtain an explicit expression for the $\beta$ parameter. 
When including nonlinearity, we find that a local, or nonlocal, intensity dependent refractive index perturbation 
forces the self-trapped beam to acquire a shape correspondent
to the so-called maximally localized states in QG.
Previously investigated QG models, as the generalized harmonic oscillator,
directly describe nonlinear waves in the presence of an highly nonlocal nonlinearity.
In addition, the natural discreteness of quantized geometry enables to map
a continuous equation, as the one describing non-paraxial optical solitons, to an exact discrete model.
The  techniques derived from the GUP framework allow to predict the existence of sub-wavelength 
optical beams for both focusing and defocusing nonlinearities, and open the way to the direct experimental tests
 of the generalized quantum mechanics supposed to be valid at the Planck scale.

\noindent {\it Projective GUP ---} The simplest generalized Schr\"odinger equation (GSE) sustaining a GUP is \cite{Das08}
 \begin{equation}
 \imath\hbar \partial_t \psi=\frac{\hat{p}^2}{2m}\psi+\frac{\beta}{m} \hat{p}^4\psi\text{,}
 \label{GUPeq}
 \end{equation}
 with $\hat p=-\imath\hbar\nabla$ the standard momentum operator. 
This equation arises when projecting the Helmholtz equation for the electromagnetic 
field $\mathcal{E}$ with wavenumber $k$ in the forward direction $z$, which is the equation
that describes light propagation beyond the paraxial approximation (see, e.g, \cite{Bulso14} and references therein):
 \begin{equation}
 i \partial_z \mathcal{E}+\sqrt{\nabla^2+k^2}\mathcal{E}=0\text{.}
\label{fwdhelm}
 \end{equation}
 By $\psi=\mathcal{E}\exp(i k z)$, Eq.(\ref{fwdhelm}) is written as
 \begin{equation}
 \imath\lambdabar \partial_z \psi= \mathcal{\hat P}_z \psi=\left[1-\sqrt{1-(-\imath\lambdabar\nabla)^2} \right]\psi
 \label{genschr}
 \end{equation}
 with $\lambda=2\pi/k$ the wavelength, and $\lambdabar=\lambda/(2\pi)$.
 $\nabla$ is the gradient with respect to the transverse coordinates $(x,y)$.
By $m c^2\equiv \hbar \omega=\hbar c /\lambdabar$, with $\omega=2\pi c/\lambda$, being $T$ the laboratory time, with $t=z/c-T$, we have
 \begin{equation}
 \imath\hbar \partial_t \psi= \hat{H} \psi= m c^2 \hat{\mathcal{P}}_z \psi\text{,}
 \label{SE}
 \end{equation}
which, for small $\lambda$, gives Eq.(\ref{GUPeq}) with
 \begin{equation}
 \beta=\frac{3}{8} \frac{1}{m^2 c^2} =\frac{3}{8}\left(\frac{\lambda}{h}\right)^2\text{.}
 \label{beta}
 \end{equation}
It is remarkable that this approach allows to obtain an expression for the $\beta$ parameter, which may be generalized to other fields.
Letting $G$ the gravitational constant,
 and $M_P=(\hbar c /G)^{1/2}$ the Planck mass, the normalized coefficient $\beta_0=(M_P^2 c^2)\beta$
 is considered, and we have
 \begin{equation}
 \beta_0=\frac{3}{8}\frac{M_P^2}{m^2}=\frac{3}{8} \frac{\hbar c}{G}\frac{c^2\lambda^2}{\hbar^2}=
 \frac{3}{8}\frac{c^3\lambda^2}{G \hbar}\text{.}
 \label{contieq1}
 \end{equation}
When applied to the photon with wavelength $\lambda=1~\mu$m, Eq.(\ref{contieq1}) gives $\beta_0=10^{55}$.  In \cite{Das08} it has been estimated $\beta_0<10^{34}$; we hence observe that, in this analogue of QG, the GUP effects are
several order of magnitudes greater than those previously estimated. This shows that GUP effects can be
directly observed in the laboratory and also furnishes a direct expression for the parameter $\beta_0$. Eq.(\ref{contieq1}) can be generalized to an higher number dimensions.

\noindent {\it Maximally localized states --- } Limiting our analysis to one dimension $x$, we have $\hat H=\hat P^2 /{2m}$, being 
 \begin{equation}
 \hat P=\sqrt{2} m c \sqrt{1-\sqrt{1-\left(\frac{\hat p}{mc}\right)^2}}\frac{\hat{p}}{|\hat{p}|}\text{,}
 \end{equation}
the generalized momentum. When $\beta\rightarrow 0$  $\hat P=\hat p \left(1+\frac{\beta}{3}\hat p\right)$.
In addition, $\hat p=\hat P\sqrt{1-(\hat P/(2 m c))^2}=\hat P (1-\beta \hat P/3)+O(\beta^2)$.
In the momentum representation, waves have finite support in the interval $p\in [-mc, mc]$, corresponding to $P\in [-\sqrt{2}mc, \sqrt{2}mc]$.
This approach leads to states with a finite support for $P$,
as those introduced by Pedram in \cite{Pedram12}.
We have (the prime denoting the derivative)
 \begin{equation}
 [\hat X, \hat P]=\imath\hbar \frac{\sqrt{1-(\hat P/ 2 m c)^2}}{1-\hat P^2/2 m^2 c^2}=\frac{\imath\hbar}{p'(\hat{P})}
 \end{equation}
 that, for $\beta\rightarrow 0$, reduces to the KMM model \cite{Kempf95}
 \begin{equation}
 [\hat X, \hat P]=\imath\hbar (1+\beta \hat P^2)\text{.}
 \end{equation}
 The GUP (\ref{GUP}) arises at the lowest order in $\beta$ from  $\Delta X \Delta P\geq |\langle [\hat X, \hat P]\rangle|$ with $\langle \hat P \rangle=0$.
This is equivalent to the original KMM proposal \cite{Kempf95}, with
  $\hat X=\hat x$,  $\hat P=\hat p + \frac{\beta}{3}\hat p^3$ being $[\hat{x},\hat{p}]=\imath\hbar$.

\noindent Being a minimal length uncertainty, the eigenstates of the position operator $\hat{X}$ are not physically realizable,
 as they correspond to $\Delta X=0$. 
In the $P-$representation $\psi(P)=\langle P| \psi \rangle$,
 and we have for the position operator
 \begin{equation}
 \hat{X}\rightarrow \frac{\imath\hbar}{p'(P)}\partial_P \psi(P)\text{.}
 \end{equation}
Letting $\hat{P}_z=(c P)^2/2$, one has
 \begin{equation}
 \imath\hbar \psi_t(P)=\frac{P^2}{2m} \psi(P)
 \end{equation}
 so that the $P-$eigenstates correspond to free particles with kinetic energy $P^2/2m$,
 which reduces to the standard $p^2/2m$ in the small momentum limit $\beta\rightarrow 0$.

 The $P-$representation is very useful to determine the maximally localized (ML) states satisfying the condition of minimal $\Delta X$.
 Following \cite{Detournay02}, ML states with $\langle X \rangle=\langle P \rangle=0$ are given by 
 \begin{equation}
 \left\{\left[\frac{\imath\hbar\partial_P}{p'(P)}\right]^2 -\mu^2\right\}\Phi^{ML}(P)=0\text{,}
\label{DGSeq}
 \end{equation}
with $\mu=\Delta X$, which leads to ($n=1,2,...$)
 \begin{equation}
 \Phi^{ML}_n(P)=\sqrt{\frac{\hbar}{mc}} \sin\left[\frac{n \pi}{2 mc }p(P)+\frac{n\pi}{2}\right]\text{.}
 \end{equation}
The maximal localization corresponds to $n=1$, with $\mu=\Delta X_{min}=\hbar \pi/2 mc=\lambda/4$.

\noindent {\it Nonlinearity ---} 
\noindent We consider the following nonlinear wave equation
 \begin{equation}
 \imath \hbar\partial_t \psi=\frac{\hat{P}^2}{2m}\psi+\int \kappa(x-x')|\psi(x')|^2 dx'\psi\text{,}
 \label{nonlinear1}
 \end{equation}
where $\kappa(x)$ is a kernel function. We consider a nonlocal 
nonlinearity (NN) \cite{Krolikowski04} because the interaction length may be much larger than the geometry quantization length. 
NN supports stable regimes, \cite{Bang02} with exact solutions in the highly \cite{Snyder97} and weakly nonlocal limits \cite{kroli00}.
Eq.(\ref{nonlinear1}) may be generalized to include vectorial effects. The strength of the nonlinearity is determined by the norm $\mathcal{N}=\int |\psi|^2 dx$.

 We first analyze the highly nonlocal (HN) case with $\kappa(x)*|\psi(x)|^2\cong \mathcal{N}\kappa(x)$ and the
 asterisk denoting the convolution integral. \cite{Folli12}
 Letting $\mathcal{N}\kappa(x)=\kappa_2 \mathcal{N}x^2/2\equiv\hbar c \Omega x^2/2$, 
with $\kappa_2$ and $\Omega^2=\kappa_2\mathcal{N}$ coefficients of the Taylor expansion
of $\kappa(x)$, we have $\hat{H}=\hat{P}^2/2m + m \Omega^2 x^2/2$. 

\noindent The SW are given by $\psi(P,t)=\phi(P)\exp(-\imath \frac{E}{\hbar} t)$ with 
 \begin{equation}
 \frac{P^2}{2m}\phi(P) -\frac{m\Omega^2 \hbar^2}{2p'(P)}\frac{\partial}{\partial P}\left[\frac{1}{p'(P)} \frac{\partial \phi}{\partial P}\right]=E \phi(P)\text{.}
 \label{nonparaxialbounds1}
 \end{equation}
 In the small momentum limit Eq.~(\ref{nonparaxialbounds1}) reduces to 
 \begin{equation}
 \frac{\partial^2 \phi}{\partial P^2}+\frac{2\beta P}{1+\beta P^2}\frac{\partial \phi}{\partial P}+\left(\frac{2E }{m\hbar^2 }-\frac{P^2}{m^2\Omega^2 \hbar^2}\right)\frac{\phi}{(1+\beta P)^2}=0\text{,}
 \label{kempfHO}
 \end{equation}
 which is the QG harmonic oscillator in \cite{Kempf95}.
We write Eq.(\ref{nonparaxialbounds1}) in dimensionless units, by $y=P/\sqrt{2}mc$ and $\epsilon=E/mc^2$
\begin{equation}
-g(y)\frac{d}{dy}\left[g(y) \frac{d}{dy}\Phi(y) \right]+\frac{y^2}{\varepsilon^2} \Phi(y)=\frac{\epsilon}{\varepsilon^2} \Phi(y) 
\label{sturm1}
\end{equation}
with $\Phi(\pm 1)=0$, $g(y)=\sqrt{1-y^2/2}/(1-y^2)$, and $\varepsilon=\hbar \Omega/2 mc^2$.

According to the Sturm-Liouville theory \cite{InceBook},
  as $g(y)>0$, Eq.(\ref{sturm1}) admits discrete solutions $\Phi_n(P)$ and a set
of positive values $\epsilon_n$ of $\epsilon$, with $n=0,1,2,...$ the number of zeros of $\Phi_n(P)$.
For $\varepsilon\rightarrow 0$, in the small nonlinearity limit, a multiple scale expansion with $Y=y/\sqrt{\varepsilon}$
and $\Phi(y)=\Phi(y/\sqrt{\epsilon})$ gives the standard quantum harmonic oscillator (QHO),
with $E=\hbar \Omega (n+1/2)$, and, for $n=0$, $\Phi(y)=exp(-\varepsilon y^2/2)$ 
and $\epsilon=\varepsilon$.
For large nonlinearity $\varepsilon \rightarrow \infty$, Eq.(\ref{sturm1}) reduces to Eq.~(\ref{DGSeq}) and the states tend to the ML: 
$\Phi(y)=\sqrt[4]{2}\cos(\pi y \sqrt{1-y^2/2}/\sqrt{2})$ with $\epsilon=\varepsilon^2 \pi^2/2$,
i.e., $E=\pi^2 (\hbar \Omega)^2/8 m c^2$.

\begin{figure}
\includegraphics[width=0.45\textwidth]{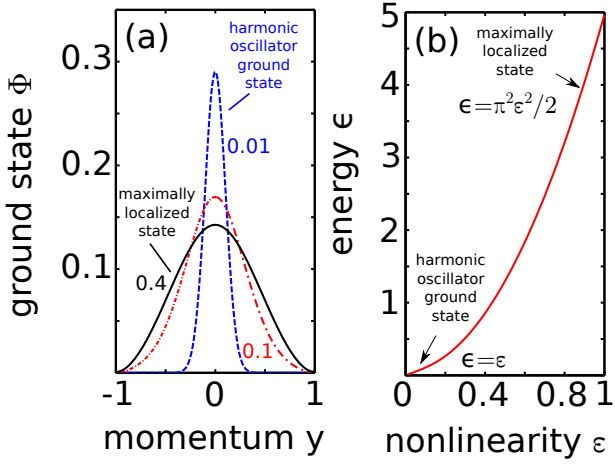}
\caption{
(Color online)
(a) Profiles of the ground state $\Phi(y)$ versus the dimensionless generalized momentum $y$ for three values of the nonlinearity $\varepsilon$;
(b) energy $\epsilon$ versus $\varepsilon$. 
\label{figurenonlocal}}
\end{figure}

\noindent To verify these limits, Eq.(\ref{sturm1}) is numerically solved 
by writing $\Phi(y)$ as a superposition of Chebyshev polynomials \cite{BoydSpectralBook}.
In Fig.~\ref{figurenonlocal}a we show the ground state solution $\Phi(y)$ for three strengths of the nonlinearity $\varepsilon$, and in Fig. \ref{figurenonlocal}b we show the eigenvalue $\epsilon=\epsilon_0$ versus $\varepsilon$.
The solutions of Eq.(\ref{sturm1}) interpolate between the QHO ground state and the ML state.

In the quasi-position space, the solution 
can be represented by cardinal functions $\phi_C=\sqrt{\lambdabar}\sin(x/\lambdabar)/\sqrt{\pi}x$,
letting $x_n=\pi\lambdabar(2n+1)/2$:
\begin{equation}
\psi(x,t)=\sum_{n=-\infty}^{\infty} \sqrt{\pi\lambdabar} \psi(x_n,t) \phi_C(x-x_n)\text{,}
\label{cardinal}
\end{equation}
with $\sqrt{\pi\lambdabar} \phi_C(x_n-x_q)=\delta_{n}^q$ the Kronecker symbol.
We use this expansion in the local case in Eq.~(\ref{nonlinear1}), with $\kappa(x)=\chi m\,c^2 \delta(x)$,with $\delta(x)$ the Dirac $\delta$ and $\chi=\pm 1$,
\begin{equation}
  \imath \hbar \partial_t \psi=\hat{H}\psi-\chi m c^2 |\psi|^2 \psi\text{.}
\label{nls}
\end{equation}
By using Eq.~(\ref{cardinal}), Eq.(\ref{nls}) is mapped to a discrete model.
By the Einstein convention, we have
\begin{equation}
\frac{\imath \hbar}{mc^2} \frac{d \psi_n}{dt}=h_{n}^{q} \psi_q- \chi |\psi_n|^2 \psi_n\text{,}
\label{discrete}
\end{equation}
being $\psi_n(t)=\psi(x_n,t)$ and
\begin{equation}
h_n^q=\left[\frac{{\hat P}^2 \psi_C(x-x_p)}{{2 m^2 c^2}}\right]_{x=x_q}=\delta_{p}^q-\frac{J_1[\pi(n-p)]}{2(n-p)}\text{.}
\end{equation}
The matrix $h_n^{q}$ is the cardinal basis representation of the operator $\hat{P}^2/(2m^2 c^2)$ and is positive-definite.
We also have, letting $\psi^n\equiv\psi_n$, 
\begin{equation}
\mathcal{N}=\pi\lambdabar\sum_{n=-\infty}^{\infty}  |\psi_n|^2=\pi \lambdabar \psi_n (\psi^n)^* \text{.}
\end{equation}
Bound states of (\ref{discrete}) are $\psi_n=\phi_n \exp(-\imath E t/\hbar)$ with
\begin{equation}
h_n^q\phi_q- \chi \phi^3_n=\frac{E}{mc^2}\phi_n\text{.}
\label{bound_state_1}
\end{equation}
To study the ground state solution of Eq.(\ref{bound_state_1}), 
we adopt a perturbation expansion in $1/\mathcal{N}$, with the ML as leading order,
\begin{equation}
\phi_n^{ML}=\sqrt{\frac{\mathcal{N}}{\lambda}} \left(\delta_{n}^0+\delta_{n}^{-1}\right)\text{.}
\end{equation}
For $\mathcal{N}\rightarrow\infty$ the solution is written as
\begin{equation}
\phi_n=\phi_n^{ML}+\frac{1}{\mathcal{N}}\phi_n^{(1)}+o(\mathcal{N}^{-1})
\label{bound_state_asymptotic}
\end{equation}
and $E/mc^2\equiv \epsilon=\epsilon^{ML}+\mathcal{N}\epsilon^{(1)}$. 
By Eq.(\ref{bound_state_1}), 
\begin{equation}
\phi_n^{(1)}=(1-\delta_n\ezero-\delta_{n}\emone)\frac{\chi\sqrt{\mathcal{N}\lambda}}{2}\left\{\frac{J_1(\pi n)}{n}+\frac{J_1\left[(n+1)\pi\right]}{n+1}\right\}\text{,}
\end{equation}
$\epsilon^{(1)}=-\chi/\lambda$, and 
\begin{equation}
\epsilon=\frac{E}{mc^2}=h\szero\ezero+h\emone\szero-\frac{\chi \mathcal{N}}{\lambda}=1-\frac{\pi}{4}-\frac{J_1(\pi)}{2}-\frac{\chi \mathcal{N}}{\lambda}\text{,}
\label{localenergy}
\end{equation}
being $h\szero\ezero=1-\pi/2>0$, and $h\szero\emone=-J_1(\pi)/2<0$.
The SW tends to ML for $\chi=\pm 1$, i.e.,
$\langle \phi^{ML} | \phi \rangle=1+o(\mathcal{N}\emone)$. 

\begin{figure}
\includegraphics[width=0.45\textwidth]{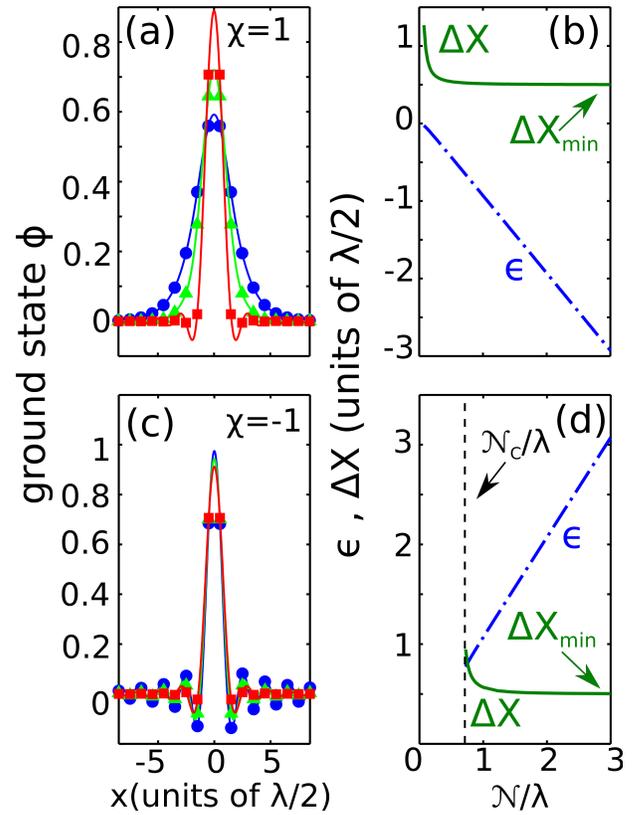}
\caption{
(Color online) (a) SW profiles scaled to unitary norm for $\chi=1$, calculated in the cardinal basis after Eq.(\ref{cardinal}) and the numerical solution of Eq.~(\ref{bound_state_1}),
for $\mathcal{N}/\lambda=0.075$ (circles), $\mathcal{N}/\lambda=0.134$ (triangles), $\mathcal{N}/\lambda=3.00$ (squares); (b) $\epsilon$ and $\Delta X$ when increasing the strength of nonlinearity ; (c) as in (a) for $\chi=-1$ and 
$\mathcal{N}/\lambda=0.85$ (circles), $\mathcal{N}/\lambda=1$ (triangles),
$\mathcal{N}/\lambda=3$ (squares); (d) as in (b) for $\chi=-1$.
\label{figurelocal}}
\end{figure}

\noindent To validate this analysis, we numerically solve Eq.(\ref{bound_state_1}) by a Newton-Raphson algorithm.
We first consider the case $\chi=1$ and we show in Fig.~ \ref{figurelocal}a the ground state expressed in terms of the
discrete values $\phi_n$ and after Eq.(\ref{cardinal}).
When increasing $\mathcal{N}$ the wavefunction tends to a ML state, 
with only two non-vanishing coefficients $\phi\szero$ and $\phi\smone$ corresponding to 
$x\szero=\pi\lambdabar=\lambda/2$ and $x\smone=-\pi\lambdabar=-\lambda/2$.
We show in Fig.~\ref{figurelocal}b the eigenvalue $\epsilon$ which, for large $\mathcal{N}$, tends to Eq.(\ref{localenergy}). 
We also show in  Fig.\ref{figurelocal}b the calculated 
$\Delta X/\pi \lambdabar=[\sum_{q}(2 q+1)\phi_q^2/4\sum \phi_q^2]^{1/2}$, which tends to the minimal uncertainty 
$\Delta X_{min}/\pi\lambdabar=1/2$ for large nonlinearity. 
Similar results are found in the case $\chi=-1$, as shown in Fig.\ref{figurelocal}c,d; following Eq.(\ref{localenergy}), the energy $\epsilon$ is positive and the algorithm converges to a localized SW for $\mathcal{N}>\mathcal{N}_C\cong 0.6\lambda$.

\noindent We investigate the local SW stability by linearizing Eq.(\ref{discrete}) by
\begin{equation}
\psi_n=\left(\phi_n+\eta_n\right) \exp{\left(-\imath E t/\hbar\right)}\text{,}
\end{equation}
which gives
\begin{equation}
  \frac{\imath \hbar}{m c^2}\frac{d\eta_n}{dt}=h_n^q \eta_q-\epsilon \eta_q+\chi \phi_n^2 (\eta_n +2 \eta_n^*)\text{.}
\label{linearstability}
\end{equation}
We numerically solve Eq.(\ref{linearstability}) for exponentially diverging solutions. 
$\phi_n$ for $\chi=-1$ is always stable.
For $\chi=1$ we find a real-valued unstable eigenvalue $\alpha$ growing with $\mathcal{N}$. 
This is analytically verified by $\eta_n=\eta\szero \delta_n\ezero+\eta\smone\delta_n\emone$ and
$\eta_\pm=\eta\szero\pm\eta\smone$, which gives 
for $\mathcal{N}\rightarrow\infty$ and $\phi_n\cong \phi_n^{ML}$,
\begin{equation}
\imath \frac{d\eta_{\pm}}{dt}=(h\szero\ezero\pm h\szero\emone) \eta_{\pm}-\epsilon \eta_\pm+\frac{\chi \mathcal{N}}{\lambda} (\eta_\pm +2 \eta_\pm^*)\text{,}
\label{linearstability1}
\end{equation}
Letting $\eta_{\pm}=\hat\eta_{\pm}\exp(\alpha_\pm m c^2 t/\hbar)$, we have $\eta_+=0$, 
and 
\begin{equation}
\alpha_-^2=4 |h\szero\emone| \left[h\szero\emone +\frac{\chi \mathcal{N}}{\lambda} \right]\text{.}
\label{unstable}
\end{equation}
For $\chi=1$, Eq.(\ref{unstable}) gives an unstable mode $\alpha_-^2>0$
with growth rate, for large $\mathcal{N}$,
$|\alpha_-|=\left[2 J_1(\pi) \mathcal{N}/\lambda\right]^{1/2}$
in agreement with the numerical solutions of Eq.~(\ref{linearstability}).
This mode exists for $\mathcal{N}/\lambda>\mathcal{N}_{th}/\lambda=J_1(\pi)/2 \cong 0.1$.
For $\chi=-1$, $\alpha_-^2$ is negative, and the mode is stable and corresponds to an energy shift $|\alpha_-|$.
As the energy $\epsilon\pm |\alpha_-|$ must be positive this gives the existence boundary
$\mathcal{N}/\lambda>\mathcal{N}_C/\lambda\cong 2 J_1(\pi)\cong 0.6$.

\noindent{\it Discussion---} The study here reported is based on a one-dimensional (1D) model and correspondingly neglects vectorial effects, which do not occur for 1D linearly polarized beams. For focusing and defocusing nonlinearities, in the limits of highly nonlocal and local nonlinearity, non-paraxial self-trapped beams can be predicted by the analogy with QG. 
During evolution, nonlinearity may enhance, or reduce, the level of localization,
and the particular regime in terms of the amount of nonlocality. Further analysis is needed to determine if the ML states may be spontaneously generated by paraxial spatially extended beams. On the contrary, the predicted QG-driven SW are expected to be accessible when the initial spatial extension is comparable with the wavelength. It may also happen that the wavepacket periodically evolves back-and-forth different nonlocal regimes. This corresponds to periodical QG-like effects when the size of the beam reaches the equivalent of the Planck scale.

The extension to two-dimensional (2D) wavepackets requires additional mathematical concepts, as reported in future work. The 1D configuration, in optics, can be either achieved by considering elliptical beams, or in a guiding slab geometry. In the former case, transverse instabilities may occur that break the 1D regime (as, e.g., investigated in \cite{Folli12PRL}). However, these instabilities are absent 
for a defocusing nonlinearity, as in thermal liquids, or require very high power levels and a shaping of the input beam (e.g., to impose an unstable periodical modulation) to be relevant. Vectorial effects are specifically absent when considering electrostrictive nonlinearities \cite{BoydBook}, or thermal focusing \cite{Gentilini:14} and defocusing phenomena \cite{Gentilini2013},
which are also known to be highly nonlocal. Nonlocality, in turn, is known to filter out unstable mechanisms.\cite{Krolikowski04}

Light absorption sustains thermal effects but may alter the nonlinear dynamics. However, as specifically true in the non-paraxial regime with tightly focused wavepackets, the absorption length $L_{abs}$ can be much larger than the nonlinear and diffraction lengths $L_d$. This fact enables to reach highly nonlinear regimes (as in \cite{Gentilini2013}) with negligible role of loss. For example, in a thermal liquid \cite{Gentilini2013} with refractive index $n_0\cong 1.3$, and $L_{abs}\cong 2$~mm much longer than $L_d$ for an elliptical beam focused in one transverse direction $x$ at the wavelength scale. For $\lambda=532~$nm, $L_d\cong 4~\mu$m$<<L_{abs}$, and the input beam can be generated by high numerical aperture objectives. Due to the very pronounced nonlinear response of thermal liquids,\cite{Gentilini2013} the QG-SW are expected at power levels of the order of $100~m$W.

\noindent{\it Conclusions ---} In this manuscript, we have studied nonlinear waves in a model based 
on a generalized Schr\"odinger equation implying a modified uncertainty principle with
minimal length uncertainty and maximal momentum; the model is obtained by projecting
in a forward spatial direction the Helmholtz equation.
We have shown that, for local and nonlocal responses, nonlinearity forces
the solitary wavepackets to reach the maximal localization.
The reported results show that some of the consequences of the generalization of quantum mechanics emerging from the physics at the Planck scale can be nowadays tested in the laboratory by using optics and photonics.
Indeed, the theoretical methods used in the generalized quantum mechanics are naturally suited for studying nonlinear optics beyond the
paraxial approximation, and may also be extended to tackle the case of ultra-short pulses without the
slowly varying envelope approximation, or discrete systems.\cite{Longhi2011,Faccio2012b,Biancalana2014,Carusotto2014}
These analyses may open the road to the first applications of quantum gravity physics,
as in microscopy, spectroscopy, and nanophotonics, as well as to the use of
quantum linear and nonlinear optics for the simulation of the physics at the Planck scale.

We acknowledge support from the Humboldt foundation and the CINECA award under the ISCRA initiative, for the availability of high performance computing.

%

\end{document}